# Landau Damping of Beam Instabilities by Electron Lenses

*V. Shiltsev, Y. Alexahin, A.Burov, and A. Valishev*

*Fermi National Accelerator Laboratory, PO Box 500, Batavia, IL 60510, USA*

*Abstract*

Modern and future particle accelerators employ increasingly higher intensity and brighter beams of charged particles and become operationally limited by coherent beam instabilities. Usual methods to control the instabilities, such as octupole magnets, beam feedback dampers and use of chromatic effects, become less effective and insufficient. We show that, in contrast, Lorentz forces of a low-energy, a magnetically stabilized electron beam, or "electron lens", easily introduces transverse nonlinear focusing sufficient for Landau damping of transverse beam instabilities in accelerators. It is also important that, unlike other nonlinear elements, the electron lens provides the frequency spread mainly at the beam core, thus allowing much higher frequency spread without lifetime degradation. For the parameters of the Future Circular Collider, a single conventional electron lens a few meters long would provide stabilization superior to tens of thousands of superconducting octupole magnets.

*Introduction.* - Collective instabilities of charged particle beams set important limitations on the beam intensity [1, 2, 3]. In general, the instability is always driven by a certain agent that, first, responds to the beam collective perturbation, and, second, acts back on it. Such responses can occur through beam-induced electromagnetic wake-fields [4], interaction with accumulated residual ions or electron clouds [3, 5].

Suppression of the collective instabilities is typically achieved by a joint action of feedback systems and Landau damping [6, 7, 8]. For multi-bunch beams, such feedbacks usually suppress the most unstable coupled-bunch and beam-beam modes. However, having limited bandwidths, these dampers are normally inefficient for the intra-bunch modes and Landau damping is needed for their suppression. To make it possible, the spectrum of incoherent, or individual particle frequencies must overlap with frequencies of the unstable collective modes, thus allowing absorption of the collective energy by the resonant particles. The frequency spread can be generated by non-linear focusing forces, such as those due to the space charge of an opposite colliding beam in colliders, or by non-linear - usually, octupole - magnets. The first option is not available at one-beam facilities, but even in the colliders, it does not exist at injection and during the acceleration ramp, where the beams do not yet collide. Thus far, commonly used are octupole magnets with the transverse magnetic fields on beam's axis of $B_x + iB_y = O_3(x + iy)^3$, which generate the transverse, or betatron, frequency shifts proportional to the square of particles' amplitudes [7]. For higher energy $E$ of the accelerated particles, the octupoles become less and less effective: the corresponding frequency spread scales as $1/E^2$ due to increasing rigidity and smaller size of the beam, while the instability growth rates scale only as $1/E$, since the transverse beam size is not important for them. As a consequence, one needs to increase the strength of these magnets accordingly. For example, in the Tevatron proton-antiproton collider, with $E \approx 1\ TeV$, there were 35 superconducting octupole magnets installed in 1 m long package cryostats and operated with up to 50 A current [9], while in the 7 TeV LHC, 336 superconducting octupole magnets, each about 0.32 m long, operate at the maximum current of 500 A [10] – and even that is not always sufficient to maintain the beam stability above certain proton bunch intensities. The anticipated $50\ TeV$ beam energy in the proton-proton Future Circular Collider (FCC-pp, [11]) would require a further factor of more than 60 in integrated octupole strength [12], which makes stabilization by octupoles greatly impractical.

Another very serious concern is that at their maximum strength, the octupoles induce significant non-linear fields and dangerous betatron frequency shifts for the larger amplitude particles, destabilizing their dynamics. This leads to increased rate of particle losses, and therefore, higher radiation load [13].

To provide a sufficient spread of the betatron frequencies without beam lifetime degradation, we propose the use of an electron lens – a high brightness low energy electron beam system [14, 15]. In this Letter, we calculate the accelerator beam coherent stability diagrams for various sizes of the electron beam, simulate numerically the effect of the electron lenses on incoherent particle dynamics and compare it with the case of octupoles. Major parameters of the electron lens devices for effective suppression of coherent instabilities are presented as examples for the LHC and for the FCC.

*Stability diagrams with electron lenses.* - The Lorenz force acting on an ultra-relativistic proton from a low energy electron beam with velocity $\beta_e c$ and current density distribution $j_e(r)$,

$$e(E_r + B_\theta) = \frac{4\pi e(1+\beta_e)}{\beta_e c}\frac{1}{r}\int_0^r j_e(r')r'dr', \qquad (1)$$

is diminishing at large radius $r$ as $\sim 1/r$; therefore, outside of the electron beam, the corresponding betatron frequency shifts $\delta\omega_{x,y}$ drop quadratically with the proton's transverse amplitudes $A_{x,y}$. For a round Gaussian-profile electron beam of rms transverse size $\sigma_e$, the amplitude dependent tune shift $\delta\omega_x/\omega_0 \equiv \delta\nu_x$, where $\omega_0$ is the proton revolution frequency, equal to [16]:

$$\delta\nu_x = 2\delta\nu_{max}\int_0^{1/2}\frac{I_0(\kappa_x u)-I_1(\kappa_x u)}{\exp(\kappa_x u + \kappa_y u)}I_0(\kappa_y u)du;$$

$$\kappa_{x,y} = \frac{A_{x,y}^2}{2\sigma_e^2}; \quad \delta\nu_{max} = \frac{I_e}{I_A}\frac{m_e}{m_p}\frac{\sigma_x^2}{\sigma_e^2}\frac{L_e}{4\pi\varepsilon_n}\frac{1+\beta_e}{\beta_e}.$$

(2)

Here $I_{0,1}(x)$ are the modified Bessel functions, $L_e$ is the length of the electron beam, $I_e$ is the electron current, $I_A = mc^3/e = 17\text{kA}$ is the Alfven current, $m_e$ and $m_p$ are electron and proton masses, $\varepsilon_n$ is the normalized rms emittance, or the action average, of the proton beam, $\sigma_x = \sqrt{\varepsilon_n\beta_x/\gamma}$ is the beam rms size, where $\beta_x$ is the ring beta-function at the lens location and $\gamma$ is the relativistic factor. The two transverse emittances and beam sizes at the lens position are assumed to be identical. The tune shift versus amplitude parameters $A_{x,y}/\sigma_e$ is shown in Figure 1.

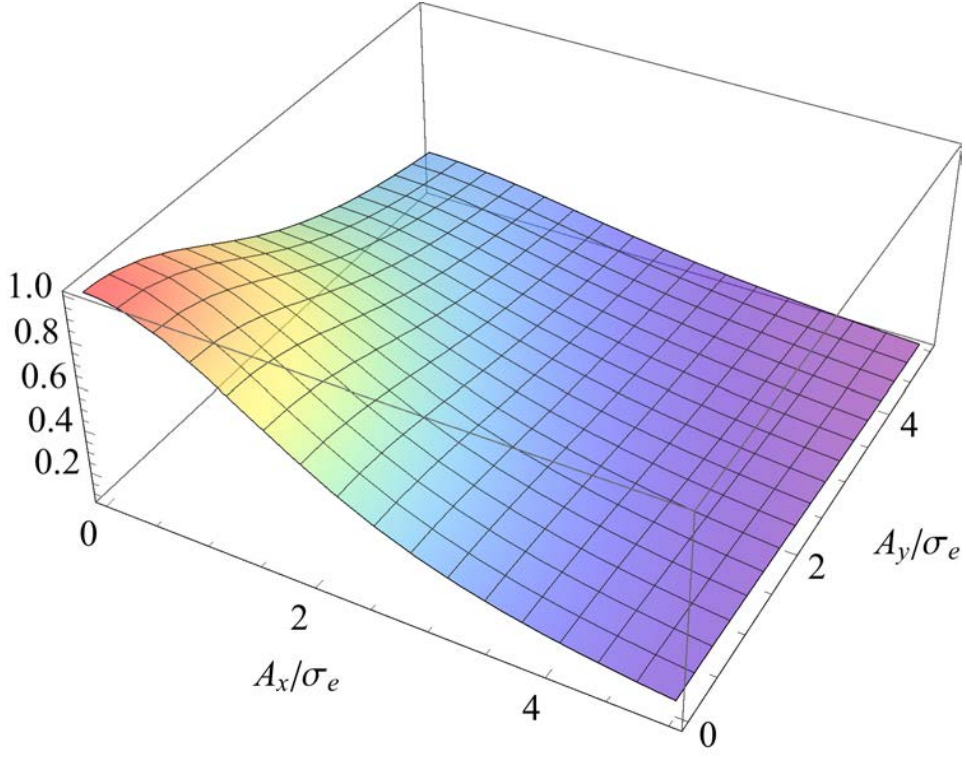

FIG. 1: The incoherent tune shift by the round electron lens, $\delta v_x / \delta v_{max}$, versus the particle transverse amplitudes, Eq. (2).

When the coherent tune shift $\Delta q$ is much smaller than the longitudinal, or the synchrotron, tune, $\Delta q \ll v_s$, which is typical for high-energy colliders with feedbacks on, the beam stability is conventionally quantified by means of the stability diagram [7]:

$$D(\Delta v) = -\left( \int \frac{J_x \partial F / \partial J_x}{\Delta v - \delta v_x + io} dJ_x dJ_y \right)^{-1}. \qquad (3)$$

Here $F$ is the normalized phase space density as a function of actions $J_{x,y}$, so that $\iint dJ_x dJ_y F(J_x, J_y) = 1$; the symbol $io$ stands for an infinitesimally small positive value in accordance with the Landau rule [6]. The function $D(\Delta\nu)$ maps the real axis in the complex plane $\Delta\nu$ onto a complex plane $D$, showing the stability thresholds for the coherent tune shifts $\Delta q$; the beam is unstable if and only if there is a collective mode whose tune shift stays above the stability diagram $D$.

In case of octupoles, the incoherent tune shifts are linear functions of the actions:

$$\begin{aligned}\delta\nu_x &= c_{xx} J_x / \varepsilon_n + c_{xy} J_y / \varepsilon_n; \\ \delta\nu_y &= c_{yx} J_x / \varepsilon_n + c_{yy} J_y / \varepsilon_n.\end{aligned} \quad (4)$$

For the LHC at 7 TeV with $\varepsilon_n = 2.5\,\mu\mathrm{m}$, its 168 Landau octupoles per beam, fed with the maximal current of 500 A, provide the nonlinearity matrix with $c_{xx} = c_{yy} = 9.5 \cdot 10^{-5};\ c_{xy} = c_{yx} = -6.7 \cdot 10^{-5}$ [8]. The corresponding stability diagram is shown in Figure 2.

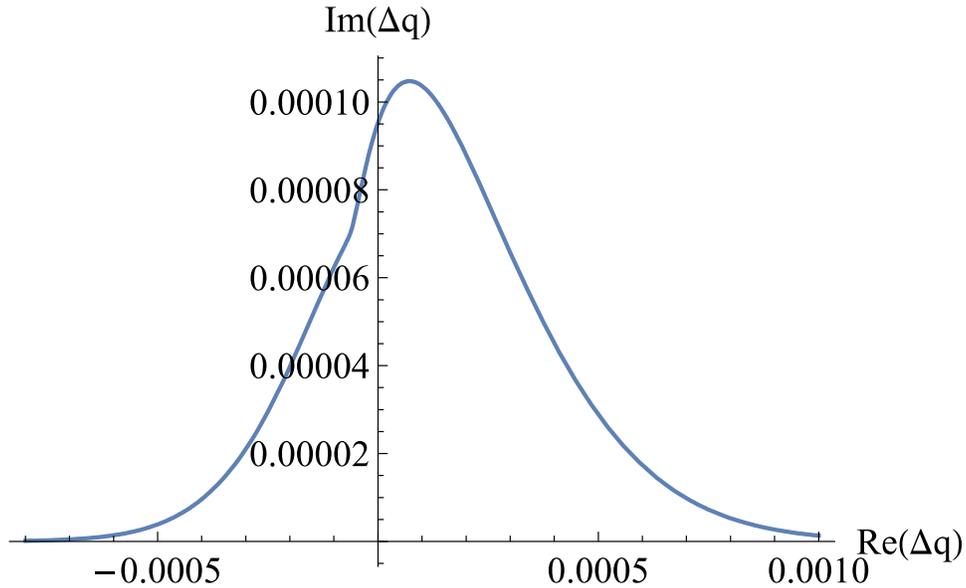

FIG. 2: Stability diagram for the 7 TeV proton beams in LHC at the maximal strength of the Landau octupoles.

For the electron lens, the stability diagram, Eq.(3), with the tune shift $\delta v_x$ given by Eq.(2), is presented in Fig. 3 for various electron beam sizes and the same current density at the center; both real and imaginary parts of the diagram are in the units of $\delta v_{max}$.

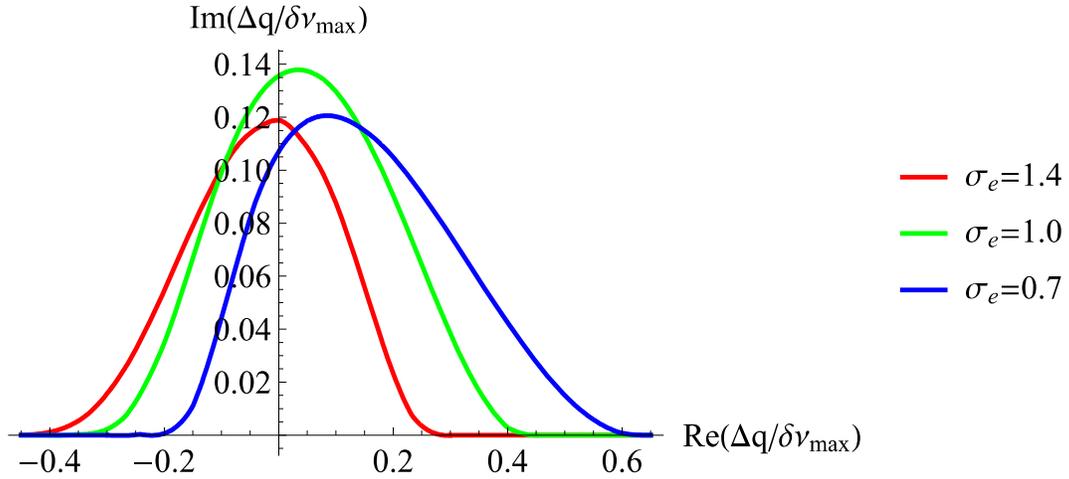

FIG. 3: Electron lens stability diagrams are presented for various electron beam sizes (noted in units of the proton beam rms size), assuming the same current density at the center.

Table I lists main parameters of the electron lens required to generate a tune spread $\delta v_{max} = 0.01$ in the LHC. For the LHC parameters, such a lens provides approximately an order of magnitude larger stability diagram than the existing Landau octupoles all operating at their maximum current of 500 A.. In the 50 TeV proton-proton Future Circular Collider, the same single lens would introduce the same tune spread $\delta \nu_{\max} = 0.01$, provided that the normalized emittance is the same and the beta-function scales as the energy, i.e. $\beta_x = 1.5 \text{km}$ at the lens location in the FCC. To make similar stability diagram for the FCC, ~20000 LHC-type octupoles would be needed. The electron system parameters listed in Table I are either modest or comparable to the electron lenses already commissioned and operational for beam-beam compensation in the Tevatron proton-antiproton collider [17, 18] and in the Relativistic Heavy Ion Collider (RHIC) [19]. Given the flexibility of the electron lenses [14], they can be effectively used for proton beam stabilization at all stages of collider operation – at injection, on the energy ramps, during the low-beta squeeze, adjustment to collisions, and, if necessary, in collisions. Moreover, the electron current can be easily regulated over short time intervals and the electron lenses can be set to operate on a subset of least stable bunches in the accelerator or even on individual bunches, as was demonstrated in the Tevatron [20]. The increased betatron frequency spread $\delta \nu$ of about 0.004-0.01 induced by the electron lenses has been demonstrated in the 980 GeV proton beam in the Tevatron [21] and in the RHIC 100 GeV polarized proton beams [22].

TABLE I: Electron beam requirements to generate the tune shift $\delta \nu_{max} = 0.01$ in the 7 TeV LHC proton beams with $\varepsilon_n = 2.5 \mu \text{m}$

| Parameter | Symbol | Value | Unit |
|---|---|---|---|
| Length | $L_e$ | 2.0 | m |
| Beta-functions at the e-lens | $\beta_{x,y}$ | 240 | m |
| Electron current | $I_e$ | 0.8 | A |
| Electron energy | $U_e$ | 10 | kV |
| $e$-beam radius in main solenoid | $\sigma_e$ | 0.28 | mm |
| Fields in main/gun solenoids | $B_m / B_g$ | 6.5/0.2 | T |
| Max. tune spread by e-lens | $\delta\nu_{max}$ | 0.01 | |

*Long-term single particle stability.-* To compare the effects of Landau damping by octupole magnets with that by the electron lenses on the long-term single particle stability, we have applied frequency map analysis (FMA) and Dynamic Aperture calculations – methods widely used to explore dynamics of Hamiltonian systems [13, 23, 24]. The phase space plot of such systems is usually a complicated mixture of periodic, quasiperiodic, and chaotic trajectories arranged in stable and unstable areas. Analysis of these trajectories and distinction between regular (periodic or quasiperiodic) and chaotic ones provides useful information on the motion features, such as working resonances, their widths, and locations in the planes of the betatron tunes and amplitudes. The FMA method is a quick tool widely used in the accelerator community for studies of particle motion stability [25, 26]. The Dynamic Aperture (DA – the area of stable long-term particle dynamics) calculation employs more computer-intensive simulations (normally hundreds of thousand or millions of turns) and is used as a figure of merit in the accelerator design and operations [27].

Figure 4 presents the simulated FMA and DA plots for the illustrative case of 7 TeV protons circulating without collisions in a focusing optics model (HL-LHC optics Version 1.0 [28]) in the presence of realistic multipole magnetic field errors in the LHC with machine chromaticity, i.e. tune derivative on the relative momentum deviation, $pdv_x/dp = 3.0$. Two Landau damping mechanisms are examined: with existing octupole magnets set to create tune spread of $\delta v = 0.01$ within the amplitudes $A_x=A_y=3.5\ \sigma_p$ (Fig. 4 a) and with a single electron lens, placed at the location IR4 of the ring such that it generates the maximal tune shift $\delta v_{max} = 0.01$ with the electron beam size matched to the proton beam size of $\sigma_p = 0.28$ mm (Fig. 4 b). The colors progressively changing from blue to red indicate the range of the betatron frequency (tune) modulation for protons from $10^{-7}$ to $10^{-3}$, respectively. The initial amplitudes $A_x$ and $A_y$ vary from 0 $\sigma_p$ (core) to 8 $\sigma_p$ (halo). Each point on the plots indicates the result of 8000 turns of tracking. The DA calculation data are shown on the same plots – the cyan lines depict the range of initial parameters beyond which particles are lost after 100,000 turns One can see a significant advantage of the dynamics with the electron lens: FMA in Fig. 4 a shows large tune variations – a clear indication of enhanced diffusion in the FMA methods – for particles with $A_{x,y} > 4\ \sigma_p$ in the case of the octupole magnets. Moreover, the particles with initial horizontal amplitude above 5 $\sigma_p$ are lost during the tracking over 8000 turns. The dynamic aperture in the case of the electron lens is significantly larger and exceeds 8 $\sigma_p$. That makes the electron lens the method of choice to provide strong Landau damping in accelerators without instigation of dangerous halo diffusion.

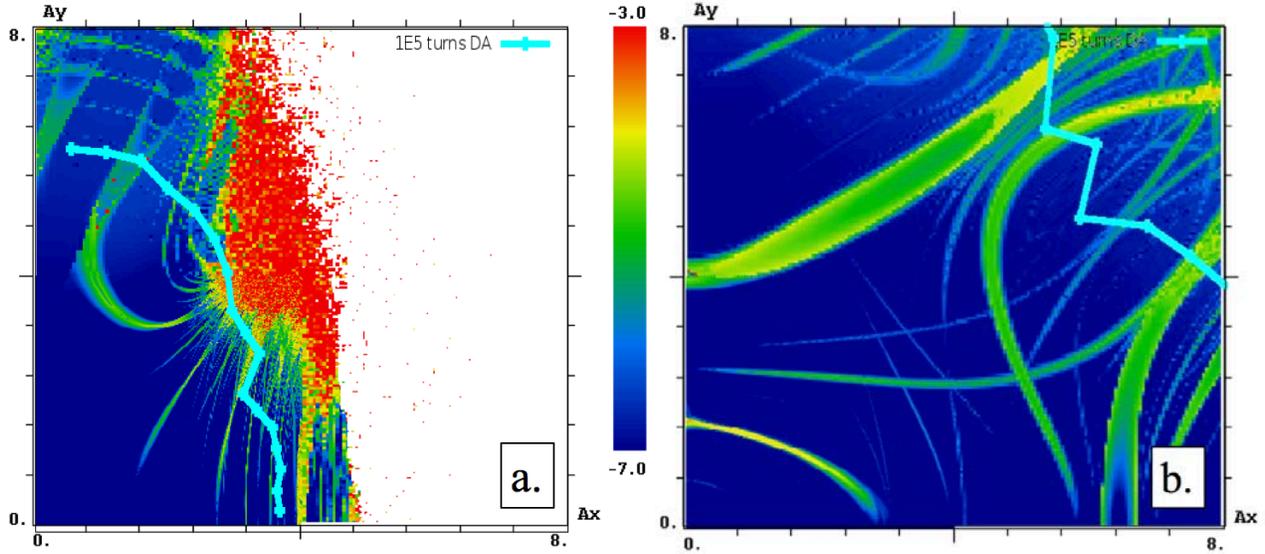

FIG. 4. Frequency Map Analysis (FMA) and Dynamic Aperture modeling of LHC proton dynamics with comparable strength Landau damping provided by octupole magnets (a) and by the electron lens (b). Horizontal and vertical axes – initial particle amplitudes $A_x$, $A_y$ in units of the rms beam size varying from 0 $\sigma_p$ (core) to 8 $\sigma_p$ (halo). Brighter colors indicate exponentially stronger tune modulation indicating resonances (see color palette). 100,000 turns DA is shown in cyan lines.

*In conclusion,* we are stressing that electron lenses are the proper Landau optical elements, since they can efficiently provide required nonlinearity where it is needed for beam stabilization, i.e. at the beam core, and do not introduce nonlinearity where it is detrimental for the lifetime, i.e. far outside the beam. Flexibility in the control of transverse electron charge distribution and fast current modulation allows the generation of the required spread of betatron frequencies by very short electron lenses with modest parameters, which have been demonstrated in the devices built so far. Landau damping by electron lenses is free of many drawbacks of other methods presently used or proposed – the lenses do not reduce the dynamic aperture and do not require numerous superconducting octupole magnets; they suppress all the unstable beam modes in contrast to available feedback systems which act only on the modes with non-zero dipole moment [8]; their efficiency will not be dependent on the bunch length as in an RF quadrupole based system [29], and corresponding single particle stability concerns due to synchro-betatron resonances will be avoided. All of this makes the Landau damping by electron lenses a unique instrument for the next generation high-current accelerators, including hadron supercolliders. Electron lenses may also be helpful in low-energy high-brightness accelerators, where Landau damping is intrinsically suppressed by a shift of single particle tunes away from the frequency of coherent oscillations [30]; a preliminary study of this issue is suggested in Ref. [31].

The technology of the electron lenses is well established and well up to the requirements of Landau damping in particle accelerators, as discussed above. Two electron lenses were built and installed in the Tevatron ring [17] at Fermilab, and two similar ones in the BNL's RHIC [22]. They employed some 10 kV Ampere-class electron beams of millimeter to submillimeter sizes with a variety of the transverse current distributions $j_e(r)$ generated at the thermionic electron gun, including Gaussian ones. The electron beams in the lenses are very stable transversely being usually immersed in a strong magnetic field - about $B_g$=0.1-0.3 T at the electron gun cathode and some $B_m$=1.0-6.5 T inside a few meters long main superconducting solenoids. The electron beam transverse alignment on the high-energy beam is done by trajectory correctors to better than a small fraction of the rms beam size $\sigma_e$. The electron lens magnetic system adiabatically compresses the electron-beam cross-section area in the interaction region by the factor of $B_m/B_g \approx 10$ (variable from 2 to 60), proportionally increasing the current density $j_e$ of the electron beam in the interaction region compared to its value on the gun cathode, usually of about 2-10 A/cm$^2$. In-depth experimental studies of Landau damping with electron lenses are being planned at Fermilab's IOTA ring [32].

We would like to thank B.Brown, E.Gianfelice-Wendt, V.Lebedev and E.Prebys for many useful comments. Fermilab is operated by Fermi Research Alliance, LLC under Contract No. DE-AC02-07CH11359 with the United States Department of Energy.